\documentclass[aip, apl, reprint, superscriptaddress]{revtex4-1}
\usepackage[english]{babel}
\usepackage{amsmath,amsthm}
\usepackage{amsfonts}
\usepackage{xspace}
\usepackage{bm}
\usepackage{booktabs}
\usepackage{graphicx}
\usepackage{dcolumn}
\usepackage[mathlines]{lineno}
\usepackage[colorlinks]{hyperref}  
\usepackage{gensymb}
\usepackage{epstopdf}
\usepackage{newfloat}
\usepackage{color,soul}
\DeclareFloatingEnvironment[name={Figure S}]{suppfigure}
\DeclareFloatingEnvironment[name={S}]{suppEquation}

\newcommand*{\citen}[1]{%
  \begingroup
    \romannumeral-`\x 
    \setcitestyle{numbers}%
    \cite{#1}%
  \endgroup   
}

\newcommand{\muN}{\ensuremath{\mu_\mathrm{0}}\xspace}
\newcommand{\aeff}{\ensuremath{\alpha_\mathrm{eff}}\xspace}

\newcommand{\Wodd}{\ensuremath{W_\mathrm{odd}}\xspace}
\newcommand{\Idc}{\ensuremath{I_\mathrm{dc}}\xspace}
\newcommand{\Jdc}{\ensuremath{J_\mathrm{dc}}\xspace}
\newcommand{\tPt}{\ensuremath{t_\mathrm{Pt}}\xspace}
\newcommand{\tYIG}{\ensuremath{t_\mathrm{YIG}}\xspace}
\newcommand{\Ms}{\ensuremath{M_\mathrm{s}}\xspace}
\newcommand{\Meff}{\ensuremath{M_\mathrm{eff}}\xspace}

\newcommand{\rPt}{\ensuremath{\rho_\mathrm{Pt}}\xspace}
\newcommand{\lPt}{\ensuremath{\lambda_\mathrm{Pt}}\xspace}
\newcommand{\SHAeff}{\ensuremath{\theta_\mathrm{eff}}\xspace}
\newcommand{\SHAeffSOT}{\ensuremath{\theta^\mathrm{SOT}_\mathrm{eff}}\xspace}

\newcommand{\SHAPt}{\ensuremath{\theta_\mathrm{Pt}}\xspace}
\newcommand{\Geff}{\ensuremath{G_{\uparrow\downarrow}^{\mathrm{eff}}}\xspace}
\newcommand{\Vish}{\ensuremath{V_{\mathrm{ISH}}}\xspace}

\begin{document}

\title{Spin-orbit torque and spin pumping in YIG/Pt with interfacial insertion layers}%

\author{Satoru Emori}%
\email{
semori@vt.edu
}
\affiliation{ 
Department of Electrical and Computer Engineering, Northeastern University, Boston, MA 02115, USA
}
\affiliation{ 
Department of Physics, Virginia Tech, Blacksburg, VA 24061, USA
}%
\author{Alexei Matyushov}%
\affiliation{ 
Department of Electrical and Computer Engineering, Northeastern University, Boston, MA 02115, USA
}%
\affiliation{ 
Department of Physics, Northeastern University, Boston, MA 02115, USA
}%
\author{Hyung-Min Jeon}%
\affiliation{ 
Materials and Manufacturing Directorate, Air Force Research Laboratory, Wright-Patterson AFB, OH 45433, USA
}%
\affiliation{ 
Department of Electrical Engineering, Wright State University, Dayton, OH 45435, USA
}%
\author{Christopher J. Babroski}%
\affiliation{ 
Department of Electrical and Computer Engineering, Northeastern University, Boston, MA 02115, USA
}%
\author{Tianxiang Nan}%
\affiliation{ 
Department of Electrical and Computer Engineering, Northeastern University, Boston, MA 02115, USA
}%
\affiliation{ 
Department of Materials Science and Engineering, University of Wisconsin-Madison, WI 53706, USA
}%
\author{Amine M. Belkessam}%
\affiliation{ 
Department of Electrical and Computer Engineering, Northeastern University, Boston, MA 02115, USA
}%
\author{John G. Jones}%
\affiliation{ 
Materials and Manufacturing Directorate, Air Force Research Laboratory, Wright-Patterson AFB, OH 45433, USA
}%
\author{Michael E. McConney}%
\affiliation{ 
Materials and Manufacturing Directorate, Air Force Research Laboratory, Wright-Patterson AFB, OH 45433, USA
}%
\author{Gail J. Brown}%
\affiliation{ 
Materials and Manufacturing Directorate, Air Force Research Laboratory, Wright-Patterson AFB, OH 45433, USA
}%
\author{Brandon M. Howe}%
\affiliation{ 
Materials and Manufacturing Directorate, Air Force Research Laboratory, Wright-Patterson AFB, OH 45433, USA
}%

\author{Nian X. Sun}%
\affiliation{ 
Department of Electrical and Computer Engineering, Northeastern University, Boston, MA 02115, USA
}%
\date{April 27, 2018}

\begin{abstract}
We experimentally investigate spin-orbit torque and spin pumping in Y$_3$Fe$_5$O$_{12}$(YIG)/Pt bilayers with ultrathin insertion layers at the interface.
An insertion layer of Cu suppresses both spin-orbit torque and spin pumping, whereas an insertion layer of Ni$_{80}$Fe$_{20}$ (permalloy, Py) enhances them, in a quantitatively consistent manner with the reciprocity of the two spin transmission processes. 
However, we observe a large enhancement of Gilbert damping with the insertion of Py that cannot be accounted for solely by spin pumping, suggesting significant spin-memory loss due to the interfacial magnetic layer. 
Our findings indicate that the magnetization at the YIG-metal interface strongly influences the transmission and depolarization of pure spin current. 
\end{abstract}
\maketitle

The transmission of pure spin current between a magnetic insulator and a normal metal is a crucial aspect of emerging insulator spintronic devices~\cite{Chumak2015, Hoffmann2015a}. 
Yttrium iron garnet (Y$_3$Fe$_5$O$_{12}$, YIG) is an especially promising magnetic insulator because of its exceptionally low Gilbert damping that allows for efficient excitation of magnetization dynamics~\cite{Howe2015, Chang2014, DAllivyKelly2013}.  
This magnetic damping can be modified by spin-orbit torque~\cite{Ando2008a, Liu2011} in thin-film YIG due to absorption of pure spin current~\cite{Wang2011a, Schreier2015, Sklenar2015a, Collet2016, Safranski2017}, which is generated from an electric current in the adjacent metal (e.g., Pt) through the spin-Hall effect~\cite{Sinova2015}. 
In the reciprocal process of spin pumping~\cite{Tserkovnyak2002, Boone2013}, coherent magnetization dynamics in YIG injects a pure spin current into the metal layer, which can be detected through an enhancement in Gilbert damping~\cite{Heinrich2011, Burrowes2012, Sun2013a} or a voltage peak due to the inverse spin-Hall effect~\cite{Weiler2013, Hahn2013a, Jungfleisch2013, Du2014d, Haertinger2015, Zhou2016, Holanda2017}. 
The reciprocity of spin-orbit torque and spin pumping is theoretically well established~\cite{Brataas2012a}.
However, while prior reports have shown that various modifications at the YIG-metal interface impact spin pumping (or, more generally, spin  transmission from the YIG to metal layer)~\cite{Burrowes2012, Sun2013a, Jungfleisch2013, Du2014d, Miao2013, Qiu2015b, Kikuchi2015}, how spin-orbit torque (i.e., spin transmission from the metal to the YIG layer) is affected by such interfacial modifications has yet to be reported. 

In this Letter, we investigate spin-orbit torque and spin pumping in the same set of YIG/Pt samples -- with and without an ultrathin interfacial insertion layer -- by ferromagnetic resonance (FMR) in a microwave cavity. 
The two spin transmission processes are suppressed with a nonmagnetic Cu insertion layer and enhanced with a magnetic Ni$_{80}$Fe$_{20}$ (permalloy, Py) insertion layer.
We also find evidence for substantial spin-memory loss~\cite{Rojas-Sanchez2014} with the insertion of ultrathin Py. 
Our findings are consistent with the reciprocity of spin-orbit torque and spin pumping, while revealing that the magnetization at the YIG-metal interface has a significant impact on the transmission and scattering of spin current. 

Epitaxial 20-nm thick YIG films were grown on Gd$_3$Ga$_5$O$_{12}$(111) substrates by pulsed laser deposition as reported in Ref.~\citen{Howe2015}.
The YIG films were transferred through ambient atmosphere to a separate deposition system for the growth of the metal overlayers. 
The YIG samples were sonicated in acetone and ethanol and, after introduction into the deposition chamber, maintained at 250$^\circ$C at 50 mTorr O$_2$ for 30 minutes to remove water and organics on the surface.  
The metal overlayers (either Pt(5 nm), Cu(0.5 nm)/Pt(5 nm), or Py(0.5 nm)/Pt(5 nm)) were deposited by dc magnetron sputtering at room temperature, base pressure of $\lesssim$$2\times10^{-7}$ Torr, and Ar sputtering pressure of 3 mTorr. 
While the RMS surface roughness of the epitaxial YIG films is only $\lesssim$0.15 nm (consistent with Ref.~\citen{Howe2015}), the nominally 0.5-nm thick Cu and Py ``dusting'' layers may not be continuous. 
Each YIG/X/Pt sample (with X = none, Cu, or Py) was patterned into a 100-$\mu$m wide, 1.5-mm long strip by photolithography and ion milling. 
The strip was contacted by Cr/Au pads on either end by photolithography, sputter deposition, and liftoff. 
This sub-mm wide strip geometry~\cite{Emori2015} allows for the use of a cavity electron paramagnetic resonance spectrometer to measure both spin-orbit torque and spin pumping. 

\begin{figure*}[tb]
  \includegraphics [width=1.75\columnwidth] {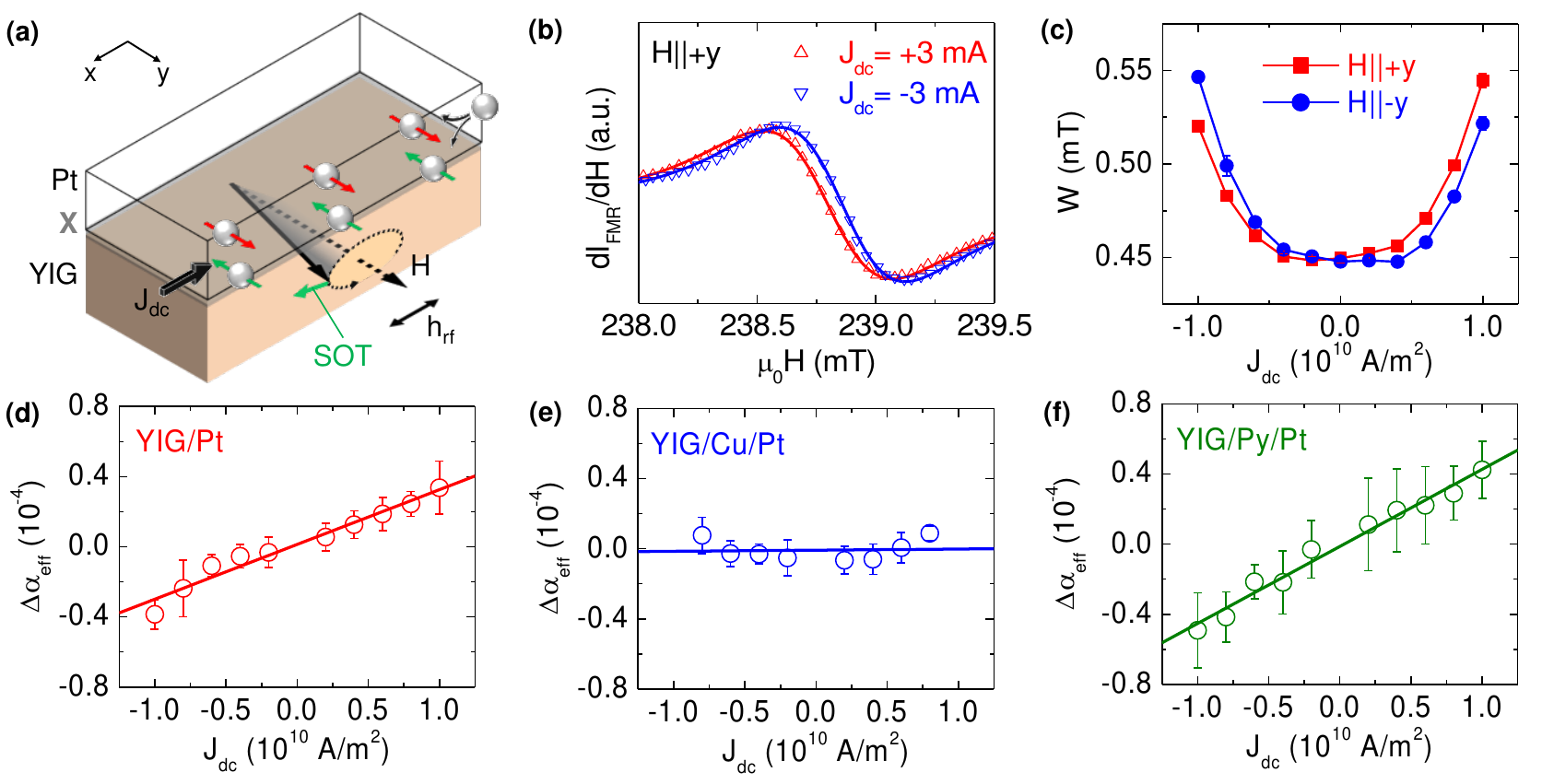}
  \centering
  \caption{\label{fig:torque}(a) Schematic of spin-orbit torque (SOT) generated by a dc current \Jdc in YIG/(X/)Pt. (b) \Jdc-induced modulation of FMR spectra in YIG/Pt. The direction of \Jdc is as defined in (a). (c) \Jdc-induced change in FMR linewidth $W$ with bias magnetic field applied along the +$y$ and -$y$ directions as defined in (a). (d-f) Change in the effective Gilbert damping parameter \aeff with \Jdc for (d) YIG/Pt, (e) YIG/Cu/Pt, and (f) YIG/Py/Pt. The lines indicate linear fits to the data.} 
\end{figure*}

We first demonstrate the transmission of spin current from the metal layer to the YIG layer through the measurement of the damping-like~\cite{Note1} spin-orbit torque.
We used a method similar to Refs.~\citen{Ando2008a, Emori2015} where the change in damping is monitored as a function of dc bias current, \Idc. 
FMR spectra were measured in a rectangular TE$_{102}$ microwave cavity with a nominal excitation power of 10 mW and several values of \Idc in the metallic layer as illustrated in Fig.~\ref{fig:torque}(a).
Each spectrum was fit with the derivative of the sum of symmetric and antisymmetric Lorentzians (e.g., Fig.~\ref{fig:torque}(b)) to extract the half-width-at-half-maximum linewidth, $W$. 

Figure~\ref{fig:torque}(c) shows the variation of $W$ with \Idc under opposite transverse external magnetic fields, $H$. 
The data contain components that are odd and even with respect to \Idc, which are due to the spin-orbit torque and Joule heating, respectively~\cite{Ando2008a, Emori2015}.
The symmetry of the spin-Hall spin-orbit torque also gives rise to a component of $W$ versus \Idc that is odd with respect to $H$ (Refs.~\citen{Ando2008a, Liu2011, Wang2011a, Emori2015}),
extracted through $\Delta\Wodd(\Idc) = \{W(\Idc, +|H|)-W(\Idc, -|H|)\}/2$. 
We can then obtain the linear change in the effective Gilbert damping parameter due to the dc spin-orbit torque, $\Delta\aeff = |\gamma|\Delta\Wodd/(2\pi f)$, where $|\gamma|/(2\pi) = 28$ GHz/T and $f = 9.55$ GHz. 

From the linear slope of $\Delta\aeff$ over the dc current density $\Jdc = \Idc/(w \tPt)$ (Fig.~\ref{fig:torque}(d)-(f)), with $w = 100$ $\mu$m and $\tPt = 5$ nm~\cite{Note2}, the effective spin-Hall angle, \SHAeff, can be quantified from~\cite{Liu2011}
\begin{equation}
\label{eq:torque}
\SHAeff = \frac{2|e|}{\hbar}\left( H+\frac{M_{eff}}{2} \right) \mu_0 \Ms \tYIG \left| \frac{\Delta\aeff}{\Jdc} \right|,
\end{equation}
where \Ms = 130 kA/m is the saturation magnetization, \Meff = 190 kA/m is the effective magnetization including the out-of-plane uniaxial anistropy field~\cite{Howe2015}, and \tYIG = 20 nm is the thickness of the YIG layer.  
By fitting the data in Fig.~\ref{fig:torque}(d) with Eq.~\ref{eq:torque}, we arrive at $\SHAeff = 0.76\pm0.05 \%$ for YIG/Pt. 

We note that \SHAeff is the product of the intrinsic spin-Hall angle of Pt, \SHAPt, and the interfacial spin current transmissivity, $T$.
Assuming that \tPt is sufficiently larger than the spin diffusion length, \lPt, the expression for \SHAeff is~\cite{Nan2015a, Pai2015a, Zhang2015e}
\begin{equation}
\label{eq:SHA}
\SHAeff = T \SHAPt \approx 2 \Geff \lPt\rPt \SHAPt,
\end{equation}
where \Geff is the effective spin-mixing conductance (which includes the spin backflow factor) and $\rPt \approx 4.0\times10^{-7}$ $\Omega$ m is the measured resistivity of the Pt layer. 
With $\lPt\rPt \approx (0.6-0.8) \times 10^{-15}$ $\Omega$m$^2$ (Refs.~\citen{Rojas-Sanchez2014, Isasa2015, Nguyen2016}), we estimate $\lPt$ to be $\approx$1.5-2~nm. 

According to Eq.~\ref{eq:SHA}, the small \SHAeff in our YIG/Pt can be attributed to a reduced $T$ (i.e., \Geff) at the YIG-Pt interface, which may be due to a residual carbon agglomeration on the YIG surface~\cite{Putter2017} that was not removed by our cleaning protocol. 
In particular, by taking $\SHAPt \approx 15-30\%$ reported from prior spin-orbit torque studies~\cite{Nan2015a, Zhang2015e, Pai2015a}, we obtain for our YIG/Pt bilayer $T\approx0.03-0.05$, or $\Geff \approx (2-5)\times10^{13}$ $\Omega^{-1}$m$^{-2}$, which is an order of magnitude lower than the typical values reported for ferromagnetic-metal/normal-metal heterostructures~\cite{Boone2013, Nan2015a, Zhang2015e, Pai2015a}, although it is comparable to prior reports on YIG/Pt~\cite{Heinrich2011, Hahn2013a}. 

\begin{figure*}[tb]
  \includegraphics [width=2.0\columnwidth] {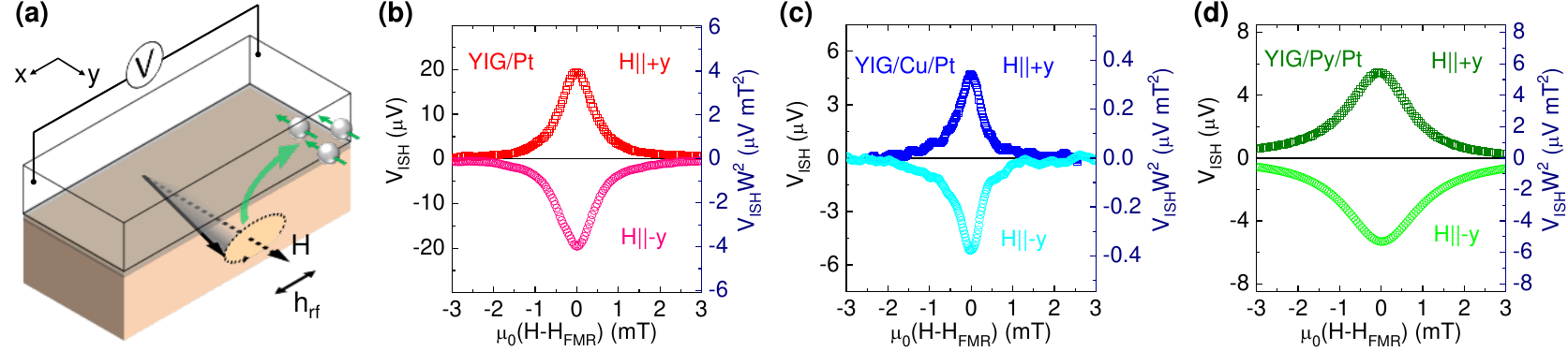}
  \centering
  \caption{\label{fig:VISH} (a) Schematic of electrically detected spin pumping in YIG/(X/)Pt. (b-d) Inverse spin-Hall voltage \Vish spectra measured for (b) YIG/Pt, (c) YIG/Cu/Pt, and (d) YIG/Py/Pt. The right vertical axis show \Vish scaled by the square of the FMR linewidth $W$, which is proportional to the transmission efficiency of spin current from YIG to Pt.} 
\end{figure*}

For YIG/Cu/Pt (Fig.~\ref{fig:torque}(e)), we do not detect a spin-orbit torque within our experimental resolution, i.e., $\SHAeff = 0.01\pm0.10\%$.
Evidently, the Cu dusting layer at the YIG-Pt interface suppresses the transmission of spin current. 
By contrast, the Py dusting layer enhances spin transmission from Pt to YIG by $\approx$40\%, with $\SHAeff = 1.08\pm0.06\%$ derived from the data in Fig.~\ref{fig:torque}(f). 
The spin-orbit torque experiment thus suggests that the nonmagnetic and magnetic insertion layers have opposite effects on spin current transmissivity (Eq.~\ref{eq:SHA}). 

In addition to spin-orbit torque, we show that the modification of the YIG-Pt interface equally impacts the reciprocal process of spin pumping.  
The same sub-mm wide YIG/X/Pt strips are measured in the setup identical to the spin-orbit torque experiment, except that the dc wire leads were connected to a nanovoltmeter, instead of a dc current source. 
As illustrated in Fig.~\ref{fig:VISH}(a), FMR in the YIG layer pumps a spin current into the Pt layer, in which the inverse spin-Hall effect converts the spin current to a charge current that is detected through a voltage peak, \Vish, coinciding with FMR. 
Figure~\ref{fig:VISH}(b)-(d) shows the \Vish spectra obtained at 10 mW of rf excitation. 
The reversal of the voltage polarity with the $H$ direction is consistent with the symmetry of the inverse spin-Hall effect.  

In the limit of \tPt sufficiently larger than \lPt, the relationship between the peak magnitude of \Vish and \SHAeff is given by~\cite{Mosendz2010a} 
\begin{equation}
\label{eq:VISH}
|\Vish| \approx \frac{h}{2|e|}\SHAeff f P \frac{L}{\tPt} \Theta^2, 
\end{equation}
where $L = 1500$ $\mu$m is the length of the sample, $P = 1.26$ is the precession ellipticity factor, and $\Theta$ is the precession cone angle. 
It should be noted that these three YIG/X/Pt samples undergo precession at different cone angles, given by $\Theta = \muN h_\mathrm{rf}/W$ (Refs.~\citen{Guan2007, Weiler2013}), since their linewidths $W$ are different. 
Due to the lack of direct calibration for the microwave field amplitude $h_\mathrm{rf}$ in our setup, the absolute magnitudes of \SHAeff cannot be determined accurately from the spin pumping experiment (Eq.~\ref{eq:VISH})~\cite{Note3}. 

Nevertheless, we can compare the \textit{relative} magnitudes of \SHAeff among the three samples.
Specifically, we scale $\Vish$ by $W^2$ ($\propto$$\Theta^{-2}$), as shown on the right vertical axis of Fig.~\ref{fig:VISH}(b)-(d), to quantify the efficiency of spin-current transmission from YIG to Pt. 
Comparing Fig.~\ref{fig:VISH}(c) with Fig.~\ref{fig:VISH}(b), the Cu dusting layer reduces the spin transmission efficiency ($\propto$$\Vish W^2$) by an order of magnitude. By contrast, comparing Fig.~\ref{fig:VISH}(d) with Fig.~\ref{fig:VISH}(b), the Py dusting layer enhnaces the transmission efficiency by $\approx$40\%. This suppression (enhancement) of spin transmission with the Cu (Py) insertion layer in the spin pumping experiment quantitatively agrees with the spin-orbit torque experiment, as summarized in Table~\ref{tab:params}. These results thus corroborate the reciprocity of the two spin-current transmission processes between YIG and Pt.

We have revealed that the ultrathin dusting layer of nonmagnetic Cu at the YIG-Pt interface suppresses spin transmission, whereas the ferromagnetic Py dusting layer enhances it. 
Our experimental results are qualitatively consistent with the first-principles calculations by Jia \textit{et al.}~\cite{Jia2011}, which report that the spin-mixing conductance at the YIG-metal interface depends on the interfacial magnetic moment density.  
With the ultrathin insertion layer of Cu (Py) decreasing (increasing) the interfacial magnetization, \Geff and hence \SHAeff decrease (increase) as described by Eq.~\ref{eq:SHA}. 
Moreover, the enhancement of spin transmission between YIG and Pt with an ultrathin ferromagnetic insertion layer, quantitatively similar to our results, has been observed in a spin-Seebeck effect experiment by Kikuchi \textit{et al.}~\cite{Kikuchi2015}.
We further note that although bulk Pt is paramagnetic, it is close to fulfilling the Stoner criterion such that the direct interface of YIG/Pt may accommodate a higher interfacial magnetic moment density~\cite{Lu2013a, Kikkawa2017} than YIG/Cu/Pt.  

The large reduction of spin-orbit torque and spin pumping with the ultrathin Cu insertion layer may seem unexpected, considering that this insertion layer is much thinner than the typical spin diffusion length of Cu ($\lambda_\mathrm{Cu}>100$ nm)~\cite{Kimura2005a}. 
Indeed, prior spin pumping experiments report only a modest decrease (by $\sim$10\%) in spin-current transmission between YIG and Pt when the Cu spacer thickness is $\approx$1 nm~\cite{Sun2013a, Du2014d}. 
However, spin pumping~\cite{Du2014d} and spin-Hall magnetoresistance~\cite{Nakayama2013} studies have shown that spin transmission decreases by an order-of-magnitude with the insertion of a Cu spacer layer, even when its thickness (e.g., $\approx$5 nm) is much smaller than $\lambda_\mathrm{Cu}$. 
Other studies also indicate large spin-memory loss at the Cu-Pt interface~\cite{Bass2007, Dolui2017}, although we do not observe a significant increase in spin dissipation (Gilbert damping) in YIG/Cu/Pt compared to uncapped YIG, as shown below. 
While further studies are required to understand the roles of the Cu spacer layer, one possibility is that spin transmission is highly sensitive to the nature of the YIG-metal interface, such as the morphology of the ultrathin Cu layer and the presence of carbon agglomeration~\cite{Putter2017}.

\begin{figure}[tb]
  \includegraphics [width=1.00\columnwidth] {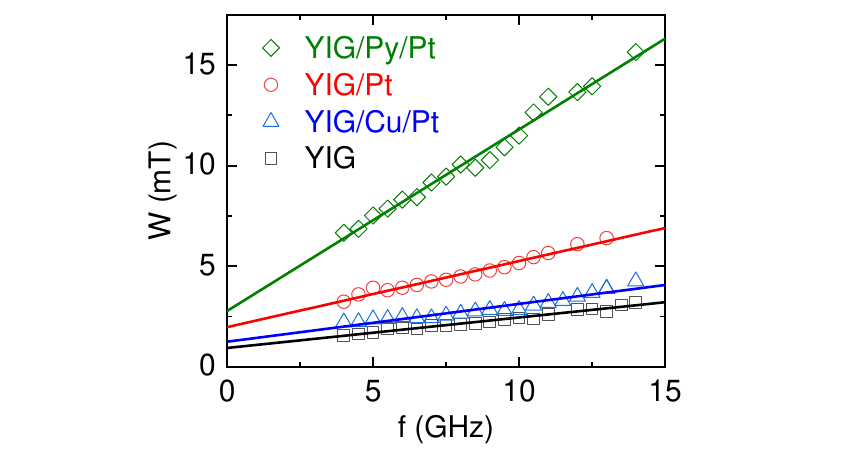}
  \centering
  \caption{\label{fig:BBFMR}
    Frequency dependence of half-width-at-half-maximum FMR linewidth, $W$.} 
\end{figure}

To gain complementary insight into interfacial spin-current transmission, we have examined the enhancement of Gilbert damping in YIG/X/Pt strips compared to uncapped YIG films. 
Fig.~\ref{fig:BBFMR} summarizes the frequency dependence of $W$, acquired with a broadband FMR setup, from which the Gilbert damping parameter, $\alpha$, is quantified. 
The averaged Gilbert damping parameter for three uncapped YIG films is $\alpha = (4.4\pm0.6)\times10^{-4}$, which is within the range reported by our earlier work~\cite{Howe2015}. 

We observe an increase in $\alpha$ for each YIG/X/Pt compared to uncapped YIG.
Assuming that the damping increase is exclusively due to spin pumping, the spin-mixing conductance is given by~\cite{Tserkovnyak2002, Boone2013},
\begin{equation}
\label{eq:GeffAlpha}
\Geff = \frac{2e^2\Ms\tYIG}{\hbar^2 |\gamma|}\Delta \alpha,
\end{equation}
where $\Delta\alpha$ (summarized in Table~\ref{tab:params}) is the difference between $\alpha$ of YIG/X/Pt and uncapped YIG. 
From Eq.~\ref{eq:GeffAlpha}, we find $\Geff = (3.3\pm0.5)\times10^{13}$ $\Omega^{-1}$m$^{-2}$ for YIG/Pt, which is in quantitative agreement with the estimated \Geff from Eq.~\ref{eq:SHA}. 
We also obtain $\Geff = (0.6\pm0.5)\times10^{13}$ $\Omega^{-1}$m$^{-2}$ for YIG/Cu/Pt, which again corroborates the one-order-of-magnitude reduction in spin transmission with the ultrathin Cu insertion layer. 
Therefore, our experimental results of spin-orbit torque (Fig.~\ref{fig:torque}), electrically detected spin pumping (Fig.~\ref{fig:VISH}), and Gilbert damping enhancement (Fig.~\ref{fig:BBFMR}) are consistent with each other for YIG/Pt and YIG/Cu/Pt. 

The Gilbert damping enhancement, $\Delta\alpha$ for YIG/Py/Pt is $\approx$4 times greater than that for YIG/Pt. 
This observation is at odds with our findings from the spin-orbit torque and spin pumping experiments, which show that \Geff (i.e., $\Delta\alpha$ according to Eq.~\ref{eq:GeffAlpha}) should be only a factor of $\approx$1.4 greater for YIG/Py/Pt compared to YIG/Pt. 
We thus estimate that only $\approx$30\% of the total $\Delta\alpha$ is due to spin pumping in YIG/Py/Pt, such that the adjusted value of \Geff is $\approx$$5\times10^{13}$ $\Omega^{-1}$m$^{-2}$. 
The remaining $\approx$70\% of $\Delta\alpha$ is likely due to spin-memory loss, i.e., spin depolarization by the ultrathin Py layer that increases the Gilbert damping but does not contribute to spin-current transmission from YIG to Pt. 
This large spin-memory loss in YIG/Py/Pt is comparable to reports on ferromagnetic-metal/Pt heterostructures~\cite{Rojas-Sanchez2014, Dolui2017, Berger2017}.

In summary, we have measured the transmission of spin current between YIG and Pt thin films, separated by an interfacial dusting layer of nonmagnetic Cu or magnetic Py, through FMR-based spin-orbit torque and spin pumping experiments. 
Spin transmission decreases by an order of magnitude when ultrathin Cu is inserted at the YIG-Pt interface and increases by $\approx$40 \% with the insertion of ultrathin Py. 
The quantitatively consistent results from the spin-orbit torque and spin pumping experiments confirm the reciprocity of these two processes. 
However, with the Py insertion layer, the Gilbert damping parameter is much larger than expected from spin pumping, suggesting substantial spin-memory loss in YIG/Py/Pt.
Our findings shed light on the roles of interfacial magnetization in the transmission and depolarization of spin current between a magnetic insulator and a normal metal. 

\textbf{Acknowledgments: }This work is funded by NSF ERC TANMS 1160504, AFRL through contract FA8650-14-C-5706, and by the W.M. Keck Foundation. 
Lithography was performed in the George J. Kostas Nanoscale Technology and Manufacturing Research Center.
We thank Peng Wei, Jean Anne Incorvia, and Jagadeesh Moodera for assistance in ion milling; and Carl Boone, Jack Brangham, Fengyuan Yang, and Branislav Nikoli{\'{c}} for helpful discussions.

\begin{table}
\caption{\label{tab:params}Essential extracted parameters - \SHAeffSOT: effective spin-Hall angle from the spin-orbit torque experiment; $\Vish W^2$: efficiency of spin transmission from the electrically detected spin pumping experiment; \Geff: effective spin-mixing conductance from the enhancement in Gilbert damping (YIG/Py/Pt adjusted to account for spin-memory loss); $\Delta\alpha$: total enhancement of the Gilbert damping parameter.} 
\begin{tabular}{l*{3}{c}r}
{}	             &YIG/Pt	&	YIG/Cu/Pt	& YIG/Py/Pt \\
\hline
\SHAeffSOT(\%) 		&$0.76\pm0.05$  		& $0.01\pm0.10$		& $1.08\pm0.06$ \\
$\Vish W^2$ ($\mu$V mT$^2$) 			&$4.0\pm0.2$  	& $0.35\pm0.02$	 &$5.6\pm0.4$ \\
 \Geff($10^{13}$ $\Omega^{-1}$m$^{-2}$)	
							&$3.3\pm0.5$ 		& $0.6\pm0.5$  	& $\approx$5	\\
$\Delta\alpha$ ($10^{-4}$) 
							&$4.8\pm0.7$  		& $0.9\pm0.7$		& $21\pm1$ \\
\end{tabular}
\end{table}


\begin{thebibliography}{10}

\bibitem{Chumak2015}
A.~V. Chumak, V.~I. Vasyuchka, A.~A. Serga, and B.~Hillebrands,
\newblock Nat. Phys. {\bf 11}, 453 (2015).

\bibitem{Hoffmann2015a}
A.~Hoffmann and S.~D. Bader,
\newblock Phys. Rev. Appl. {\bf 4}, 047001 (2015).

\bibitem{Howe2015}
B.~M. Howe, S.~Emori, H.-M. Jeon, T.~M. Oxholm, J.~G. Jones, K.~Mahalingam,
  Y.~Zhuang, N.~X. Sun, and G.~J. Brown,
\newblock IEEE Magn. Lett. {\bf 6}, 3500504 (2015).

\bibitem{Chang2014}
H.~Chang, P.~Li, W.~Zhang, T.~Liu, A.~Hoffmann, L.~Deng, and M.~Wu,
\newblock IEEE Magn. Lett. {\bf 5}, 1 (2014).

\bibitem{DAllivyKelly2013}
O.~d'Allivy Kelly, A.~Anane, R.~Bernard, J.~{Ben Youssef}, C.~Hahn, A.~H.
  Molpeceres, C.~CarreÌteÌro, E.~Jacquet, C.~Deranlot, P.~Bortolotti,
  R.~Lebourgeois, J.-C. Mage, G.~de~Loubens, O.~Klein, V.~Cros, and A.~Fert,
\newblock Appl. Phys. Lett. {\bf 103}, 082408 (2013).

\bibitem{Ando2008a}
K.~Ando, S.~Takahashi, K.~Harii, K.~Sasage, J.~Ieda, S.~Maekawa, and E.~Saitoh,
\newblock Phys. Rev. Lett. {\bf 101}, 036601 (2008).

\bibitem{Liu2011}
L.~Liu, T.~Moriyama, D.~C. Ralph, and R.~A. Buhrman,
\newblock Phys. Rev. Lett. {\bf 106}, 036601 (2011).

\bibitem{Wang2011a}
Z.~Wang, Y.~Sun, Y.-Y. Song, M.~Wu, H.~Schulthei{\ss}, J.~E. Pearson, and
  A.~Hoffmann,
\newblock Appl. Phys. Lett. {\bf 99}, 162511 (2011).

\bibitem{Schreier2015}
M.~Schreier, T.~Chiba, A.~Niedermayr, J.~Lotze, H.~Huebl, S.~Gepr{\"{a}}gs,
  S.~Takahashi, G.~E.~W. Bauer, R.~Gross, and S.~T.~B. Goennenwein,
\newblock Phys. Rev. B {\bf 92}, 144411 (2015).

\bibitem{Sklenar2015a}
J.~Sklenar, W.~Zhang, M.~B. Jungfleisch, W.~Jiang, H.~Chang, J.~E. Pearson,
  M.~Wu, J.~B. Ketterson, and A.~Hoffmann,
\newblock Phys. Rev. B {\bf 92}, 174406 (2015).

\bibitem{Collet2016}
M.~Collet, X.~de~Milly, O.~d'Allivy Kelly, V.~V. Naletov, R.~Bernard,
  P.~Bortolotti, J.~{Ben Youssef}, V.~E. Demidov, S.~O. Demokritov, J.~L.
  Prieto, M.~Mu{\~{n}}oz, V.~Cros, A.~Anane, G.~de~Loubens, and O.~Klein,
\newblock Nat. Commun. {\bf 7}, 10377 (2016).

\bibitem{Safranski2017}
C.~Safranski, I.~Barsukov, H.~K. Lee, T.~Schneider, A.~A. Jara, A.~Smith,
  H.~Chang, K.~Lenz, J.~Lindner, Y.~Tserkovnyak, M.~Wu, and I.~N. Krivorotov,
\newblock Nat. Commun. {\bf 8}, 117 (2017).

\bibitem{Sinova2015}
J.~Sinova, S.~O. Valenzuela, J.~Wunderlich, C.~H. Back, and T.~Jungwirth,
\newblock Rev. Mod. Phys. {\bf 87}, 1213 (2015).

\bibitem{Tserkovnyak2002}
Y.~Tserkovnyak, A.~Brataas, and G.~Bauer,
\newblock Phys. Rev. B {\bf 66}, 224403 (2002).

\bibitem{Boone2013}
C.~T. Boone, H.~T. Nembach, J.~M. Shaw, and T.~J. Silva,
\newblock J. Appl. Phys. {\bf 113}, 153906 (2013).

\bibitem{Heinrich2011}
B.~Heinrich, C.~Burrowes, E.~Montoya, B.~Kardasz, E.~Girt, Y.-Y. Song, Y.~Sun,
  and M.~Wu,
\newblock Phys. Rev. Lett. {\bf 107}, 066604 (2011).

\bibitem{Burrowes2012}
C.~Burrowes, B.~Heinrich, B.~Kardasz, E.~A. Montoya, E.~Girt, Y.~Sun, Y.-Y.
  Song, and M.~Wu,
\newblock Appl. Phys. Lett. {\bf 100}, 092403 (2012).

\bibitem{Sun2013a}
Y.~Sun, H.~Chang, M.~Kabatek, Y.-Y. Song, Z.~Wang, M.~Jantz, W.~Schneider,
  M.~Wu, E.~Montoya, B.~Kardasz, B.~Heinrich, S.~G.~E. te~Velthuis,
  H.~Schultheiss, and A.~Hoffmann,
\newblock Phys. Rev. Lett. {\bf 111}, 106601 (2013).

\bibitem{Weiler2013}
M.~Weiler, M.~Althammer, M.~Schreier, J.~Lotze, M.~Pernpeintner, S.~Meyer,
  H.~Huebl, R.~Gross, A.~Kamra, J.~Xiao, Y.-T. Chen, H.~Jiao, G.~E.~W. Bauer,
  and S.~T.~B. Goennenwein,
\newblock Phys. Rev. Lett. {\bf 111}, 176601 (2013).

\bibitem{Hahn2013a}
C.~Hahn, G.~de~Loubens, O.~Klein, M.~Viret, V.~V. Naletov, and J.~{Ben
  Youssef},
\newblock Phys. Rev. B {\bf 87}, 174417 (2013).

\bibitem{Jungfleisch2013}
M.~B. Jungfleisch, V.~Lauer, R.~Neb, A.~V. Chumak, and B.~Hillebrands,
\newblock Appl. Phys. Lett. {\bf 103}, 022411 (2013).

\bibitem{Du2014d}
C.~Du, H.~Wang, F.~Yang, and P.~C. Hammel,
\newblock Phys. Rev. Appl. {\bf 1}, 044004 (2014).

\bibitem{Haertinger2015}
M.~Haertinger, C.~H. Back, J.~Lotze, M.~Weiler, S.~Gepr{\"{a}}gs, H.~Huebl,
  S.~T.~B. Goennenwein, and G.~Woltersdorf,
\newblock Phys. Rev. B {\bf 92}, 054437 (2015).

\bibitem{Zhou2016}
H.~Zhou, X.~Fan, L.~Ma, L.~Cui, C.~Jia, S.~Zhou, Y.~S. Gui, C.-M. Hu, and
  D.~Xue,
\newblock Appl. Phys. Lett. {\bf 108}, 192408 (2016).

\bibitem{Holanda2017}
J.~Holanda, O.~{Alves Santos}, R.~L. Rodr{\'{i}}guez-Su{\'{a}}rez, A.~Azevedo,
  and S.~M. Rezende,
\newblock Phys. Rev. B {\bf 95}, 134432 (2017).

\bibitem{Brataas2012a}
A.~Brataas, Y.~Tserkovnyak, G.~E.~W. Bauer, and P.~J. Kelly,
\newblock {Spin pumping and spin transfer},
\newblock in {\em Spin Current}, (Oxford University Press, 2012), chap.~8, pp. 87--135.

\bibitem{Miao2013}
B.~F. Miao, S.~Y. Huang, D.~Qu, and C.~L. Chien,
\newblock Phys. Rev. Lett. {\bf 111}, 066602 (2013).

\bibitem{Qiu2015b}
Z.~Qiu, D.~Hou, K.~Uchida, and E.~Saitoh,
\newblock J. Phys. D. Appl. Phys. {\bf 48}, 164013 (2015).

\bibitem{Kikuchi2015}
D.~Kikuchi, M.~Ishida, K.~Uchida, Z.~Qiu, T.~Murakami, and E.~Saitoh,
\newblock Appl. Phys. Lett. {\bf 106}, 082401 (2015).

\bibitem{Rojas-Sanchez2014}
J.-C. Rojas-S{\'{a}}nchez, N.~Reyren, P.~Laczkowski, W.~Savero, J.-P.
  Attan{\'{e}}, C.~Deranlot, M.~Jamet, J.-M. George, L.~Vila, and
  H.~Jaffr{\`{e}}s,
\newblock Phys. Rev. Lett. {\bf 112}, 106602 (2014).

\bibitem{Emori2015}
S.~Emori, T.~Nan, T.~M. Oxholm, C.~T. Boone, J.~G. Jones, B.~M. Howe, G.~J.
  Brown, D.~E. Budil, and N.~X. Sun,
\newblock Appl. Phys. Lett. {\bf 106}, 022406 (2015).

\bibitem{Note1}
{\protect In principle, it is possible to quantify the field-like
  spin-orbit torque (combined with the Oersted field) from the linear shift of
  the resonance field \protect \ensuremath {H_\protect \mathrm {FMR}}\protect
  \xspace with \protect \ensuremath {I_\protect \mathrm {dc}}\protect \xspace .
  However, because of the large quadratic change of \protect \ensuremath
  {H_\protect \mathrm {FMR}}\protect \xspace with \protect \ensuremath
  {I_\protect \mathrm {dc}}\protect \xspace ($\approx $0.3 mT/mA$^2$) due to
  thermal effects, it is difficult to reliably extract the linear \protect
  \ensuremath {I_\protect \mathrm {dc}}\protect \xspace -induced shift of
  \protect \ensuremath {H_\protect \mathrm {FMR}}\protect \xspace . The upper
  bound of the linear \protect \ensuremath {H_\protect \mathrm {FMR}}\protect
  \xspace shift is 0.01 mT per mA of \protect \ensuremath {I_\protect \mathrm
  {dc}}\protect \xspace , about a factor of $\approx $1.5 greater than the
  expected Oersted field of $\mu _0|\protect \ensuremath {I_\protect \mathrm
  {dc}}\protect \xspace |/(2w)= 0.0063$ mT per mA, in the direction of the
  Oersted field for all samples.}

\bibitem{Note2}
For simplicity, we assume that all of the dc bias current flows through the Pt
  layer. This simplifying assumption is supported by the resistance ($R\approx
  1200$ $ \Omega $) of YIG/Pt, YIG/Cu/Pt, and YIG/Py/Pt being within a few \%
  of each other.

\bibitem{Nan2015a}
T.~Nan, S.~Emori, C.~T. Boone, X.~Wang, T.~M. Oxholm, J.~G. Jones, B.~M. Howe,
  G.~J. Brown, and N.~X. Sun,
\newblock Phys. Rev. B {\bf 91}, 214416 (2015).

\bibitem{Pai2015a}
C.-F. Pai, Y.~Ou, L.~H. Vilela-Le{\~{a}}o, D.~C. Ralph, and R.~A. Buhrman,
\newblock Phys. Rev. B {\bf 92}, 064426 (2015).

\bibitem{Zhang2015e}
W.~Zhang, W.~Han, X.~Jiang, S.-H. Yang, and S.~{S. P. Parkin},
\newblock Nat. Phys.  (2015).

\bibitem{Isasa2015}
M.~Isasa, E.~Villamor, L.~E. Hueso, M.~Gradhand, and F.~Casanova,
\newblock Phys. Rev. B {\bf 91}, 024402 (2015).

\bibitem{Nguyen2016}
M.-H. Nguyen, D.~C. Ralph, and R.~A. Buhrman,
\newblock Phys. Rev. Lett. {\bf 116}, 126601 (2016).

\bibitem{Putter2017}
S.~P{\"{u}}tter, S.~Gepr{\"{a}}gs, R.~Schlitz, M.~Althammer, A.~Erb, R.~Gross,
  and S.~T.~B. Goennenwein,
\newblock Appl. Phys. Lett. {\bf 110}, 012403 (2017).

\bibitem{Mosendz2010a}
O.~Mosendz, V.~Vlaminck, J.~E. Pearson, F.~Y. Fradin, G.~E.~W. Bauer, S.~D.
  Bader, and A.~Hoffmann,
\newblock Phys. Rev. B {\bf 82}, 214403 (2010).

\bibitem{Guan2007}
Y.~Guan, W.~Bailey, E.~Vescovo, C.-C. Kao, and D.~Arena,
\newblock J. Magn. Magn. Mater. {\bf 312}, 374 (2007).

\bibitem{Note3}
{\protect For the absolute magnitudes of \protect \ensuremath
  {\theta _\protect \mathrm {eff}}\protect \xspace to be quantitatively
  consistent between the spin-orbit torque and spin pumping experiments, the
  microwave field amplitude would be $\mu _0 h_\protect \mathrm {rf}\approx 8$
  $\mu $T.}

\bibitem{Jia2011}
X.~Jia, K.~Liu, K.~Xia, and G.~E.~W. Bauer,
\newblock Europhys. Lett. {\bf 96}, 17005 (2011).

\bibitem{Lu2013a}
Y.~M. Lu, Y.~Choi, C.~M. Ortega, X.~M. Cheng, J.~W. Cai, S.~Y. Huang, L.~Sun,
  and C.~L. Chien,
\newblock Phys. Rev. Lett. {\bf 110}, 147207 (2013).

\bibitem{Kikkawa2017}
T.~Kikkawa, M.~Suzuki, J.~Okabayashi, K.-i. Uchida, D.~Kikuchi, Z.~Qiu, and
  E.~Saitoh,
\newblock Phys. Rev. B {\bf 95}, 214416 (2017).

\bibitem{Kimura2005a}
T.~Kimura, J.~Hamrle, and Y.~Otani,
\newblock Phys. Rev. B {\bf 72}, 014461 (2005).

\bibitem{Nakayama2013}
H.~Nakayama, M.~Althammer, Y.-T. Chen, K.~Uchida, Y.~Kajiwara, D.~Kikuchi,
  T.~Ohtani, S.~Gepr{\"{a}}gs, M.~Opel, S.~Takahashi, R.~Gross, G.~E.~W. Bauer,
  S.~T.~B. Goennenwein, and E.~Saitoh,
\newblock Phys. Rev. Lett. {\bf 110}, 206601 (2013).

\bibitem{Bass2007}
J.~Bass and W.~P. Pratt,
\newblock J. Phys. Condens. Matter {\bf 19}, 183201 (2007).

\bibitem{Dolui2017}
K.~Dolui and B.~K. Nikoli{\'{c}},
\newblock Phys. Rev. B {\bf 96}, 220403 (2017).

\bibitem{Berger2017}
A.~J. Berger, E.~R.~J. Edwards, H.~T. Nembach, O.~Karis, M.~Weiler, and T.~J.
  Silva,
\newblock (2017), arXiv:1711.07654.

\end{thebibliography}
\end{document}